\documentclass[preprint]{revtex4}
\usepackage{graphicx}
\usepackage{mathtools}
\usepackage{color}
\usepackage{ulem}

\begin{document}

\title{Creation and annihilation of mobile fractional solitons in atomic chains} 

\author{Jae Whan Park$^1$}
\author{Eui Hwan Do$^{1,2}$}
\author{Jin Sung Shin$^{1,2}$}
\author{Sun Kyu Song$^1$}
\author{Oleksandr Stetsovych$^3$}
\author{Pavel Jelinek$^3$}
\author{Han Woong Yeom$^{1,2}$}
\email{yeom@postech.ac.kr}
\affiliation{$^1$Center for Artificial Low Dimensional Electronic Systems,  Institute for Basic Science (IBS),  77 Cheongam-Ro, Pohang 37673, Korea.} 
\affiliation{$^2$Department of Physics, Pohang University of Science and Technology, Pohang 37673, Korea}
\affiliation{$^3$Institute of Physics of the Czech Academy of Sciences, Cukrovarnicka 10, 18221 Prague 6, Czech Republic}


\begin{abstract}
Localized modes in one dimensional topological systems, such as Majonara modes in topological superconductors, are promising platforms for robust information processing.  
In one dimensional topological insulators, mobile topological solitons are expected but have not been fully realized yet.  
We discover fractionalized phase defects moving along trimer silicon atomic chains formed along step edges of a vicinal silicon surface.
Tunneling microscopy identifies local defects with phase shifts of 2$\pi$/3 and 4$\pi$/3 with their electronic states within the band gap and with their motions activated above 100 K. 
Theoretical calculations reveal the topological soliton origin of the phase defects with fractional charges of $\pm$2$e$/3 and $\pm$4$e$/3.
An individual soliton can be created and annihilated at a desired location by current pulse from the probe tip.  
Mobile and manipulatable topological solitons discovered here provide a new platform of robustly-protected informatics with extraordinary functionalities.
\end{abstract}

\maketitle

\section*{INTRODUCTION}

Localized topological modes, such as Majorana edge modes in topological superconductors \cite{dego13,lin17} and skyrmion excitations in magnetic materials \cite{fert17,chac18}, are attracting great interest as promising platforms for robust information processing \cite{hasa10,qi11}. 
For one dimensional (1D) topological insulators, another kind of topological local modes, solitons \cite{su79,su80,heeg88,hern20}, has been known for a long time. 
Topological solitons, which have both edge-mode and excitation characteristics, have been identified not only in spontaneous 1D insulators such as polyacetylene chains \cite{heeg88} and surface atomic chains \cite{cheo15} but also in ultracold atoms \cite{atal13,he18,lchen18}, photonic crystals \cite{krau12,verb13,posh15}, and acoustic lattices \cite{xiao15,yang16}. 
In contrast to a Majonara edge mode, solitons can move fast with topologically protected information as in the case of an unpinned skyrmion.
The technology of using solitons as the robust media of delivering information was well established in classical wave systems based on optical solitons \cite{zyss98,ouzo03}.
In quantum mechanical systems, solitons can provide even more exciting opportunities such as the multilevel information processing \cite{hata97,akhm00}, quantum entanglements \cite{szum15,pend20}, and the utilization of fractional quanta \cite{su81,schr81,rice82}. 
Among these exciting possibilities, only the multilevel information processing has been demonstrated recently by the $Z_{4}$ solitons in indium atomic chains \cite{kim17}.
However, the soliton motion is largely prohibited by pinning defects or the interchain interaction in most 1D electronic systems~\cite{kim17,lee19}, making the realization of a mobile soliton with fractionalized quanta and quantum information a long-standing challenge.   
Beyond observing the existence and the interaction of solitons, the generation and manipulation of individual solitons in electronic systems has to be demonstrated for many possible applications. 

Among various proposals in these challenges \cite{su81,schr81,rice82}, the trimer chains have been the most widely discussed with a particular focus on fractional charges.
In 1D electronic systems of trimers, solitons are endowed with fractional charges of $\pm$2$e$/3 and $\pm$4$e$/3 in contrast to integer charges of solitons in conventional dimer chains due to the spin degree of freedom \cite{su79,su80}.
That is, trimer solitons are the simplest form of fractionalized solitons in an electronic system.
Contrary to the simplicity, theoretical works reveal various exotic properties of solitons in trimer systems \cite{dros17,jin17,liu17,mart19,conz20}. 
Moreover, considering the well established ternary computing architecture \cite{conn08} and the current interest in the ternary system for low power and/or neuromorphic computing systems \cite{jeon19}, the utilization of topologically protected trimer solitons is expected to expedite exciting development in information technology.
However, no electronic system with trimer solitons has been identified yet.

In this respect, the silicon atomic chains on a vicinally-cut silicon crystal [Si(553)] has attracted our attention.
By adsorption of a proper amount of gold atoms, a regular array of step-edge silicon chains is stabilized with unsaturated dangling bonds. 
This system was found to transit into a trimer structure below about 200 K \cite{ahn05,snij06,brau18} and the existence of the phase defects were noticed with their mobility and topological nature unknown \cite{snij06,shin12,hafk20}. 
In the present work, we directly identify individual mobile solitons along these trimer atomic chains by scanning tunneling microscopy and spectroscopy (STM/S).
Two different types of solitons with fractionalized (2$\pi$/3 and 4$\pi$/3) phase shifts, respectively, are observed, which are immobile at low temperature but their motion turns on above 100 K.
Their solitonic property is unambiguously confirmed by their in-gap electronic states and their immunity for scattering.
Density functional theory (DFT) and tight-binding calculations reveal further the topological properties of these solitons and their fractionalized charges.
This is the first direct microscopic observation of a mobile soliton and a trimer soliton in an electronic system.
We also succeed to generate and annihilate a soliton on a desired location by the tunneling electron pulse from the probe tip, making the first step toward the manipulation of individual solitons.
An avenue toward exploiting mobile and robust carriers of fractional quanta is thus widely opened. 

\section*{RESULTS}

\subsection*{Direct observation of mobile phase defects}
The surface of a vicinal Si(553) crystal with an optimized coverage of Au adatoms form a well ordered array of Si and 
Au atomic chains with very narrow (1.3 nm in width) terraces (Fig. 1g) \cite{erwi10,aulb13,brau18}. Each terrace consists basically of double Au chains and a Si honeycomb chain on its topmost layer (Fig. 1d) \cite{erwi10,aulb13,brau18} (more detailed atomic structure in Supplementary Fig. 6).
What concern the present work are step-edge Si atoms with dangling bonds, which correspond to one side of the Si honeycomb chain (blue and red balls in Fig. 1c) and to the rows of bright protrusions in the STM topographs (Fig. 1b).
Its low-temperature atomic structure has long been intrigued with contradictory suggestions of a CDW insulator with a periodic lattice distortion \cite{ahn05,snij06} and an antiferromagnetic insulator with a spin ordering \cite{erwi10}.   
Very recent DFT calculations successfully found a distorted CDW structure explaining most of the experimental data \cite{brau18}.
Below the transition temperature of 200 K, the STM images exhibit a structural distortion in a high empty-state bias, namely, the alternation of bright and dim protrusions in a 3a$_0$ (a$_0$, silicon surface unit cell of 0.384 nm) periodicity (Fig. 1b), which represent a monomer and a dimer in each trimer unit cell, respectively.
As detailed below, this distorted structure is a 1D CDW state as driven by the quasi 1D metallic band of unsaturated dangling bonds of step-edge Si atoms (Fig. 2a).  

The silicon trimer chains are well known to contain extrinsic defects, which appear as missing bright protrusions in high bias STM images~\cite{snij06, shin12}. 
However, extra local features appear with bright contrast when we lower the bias closer to the Fermi energy where the 3a$_0$ periodic modulation in STM becomes weak (Fig. 1a).
A careful inspection of this extra feature back in the high bias image reveals the presence of a phase mismatch of the 3a$_0$ periodicity such as a $\times$4 (4a$_0$) or a $\times$5 (5a$_0$) ($\times$2 (2a$_0$)) unit with gradually decreasing amplitude of the 3a$_0$ protrusions (Fig. 1b and Supplementary Fig. 2).
Moreover, the hopping of the phase defect is frequently noticed by the sudden a$_0$ shift of the 3a$_0$ modulations (Figs. 1e and 1f) and its motion is even directly imaged in sequential STM images (Fig. 1g and Supplementary Movie 1).   
The enhanced contrast of the phase defects in the low bias suggests the existence of a localized in-gap state.
These observations indicate that the trimer Si chains have mobile topological solitons emerging from its 1D CDW states as revealed unambiguously below.
Note that the previous observations of the phase defects \cite{snij06,shin12,hafk20} had no means to reveal their intrinsic soliton nature.  

\subsection*{Electronic structure of Si atomic chains}
The undistorted Si step-edge chain has a strongly-1D and partially-filled electronic band due to its dangling bond electrons (Supplementary Fig. 8a). 
In the present structure model, fully relaxed within the DFT calculations (Fig. 1d)  \cite{brau18}, every third Si atom along the step edge is distorted downward to split the band with an energy gap of 0.6 eV at the Fermi level (Fig. 2a).
The band gap is due to the rehybridization of $sp{^3}$ dangling bonds into $sp{^2}$ and $p$ orbitals; the unoccupied $p$ bands around 0.2 eV from distorted Si atoms (red balls in Fig. 1d) and the occupied $sp{^2}$ bands around -0.4 and -0.7 eV from undistorted Si atoms (blue balls).
This electronic structure is consistent with the spectroscopy observation shown in Fig. 3d. 

This band structure can be described well with a much simpler 1D tight-binding model considering only the single Si zigzag chain at the step edge (yellow lines in Fig. 2a). 
The neighboring Au chains (the bands of dashed lines) affect only the fine structures of the valence bands around -0.4 $\sim$ -0.7 eV, which do not affect the following discussion (Supplementary Fig. 11). 
This 1D tight binding model is straightforwardly transformed into a trimer Su-Schrieffer-Heeger (SSH) Hamiltonian as described by three hopping amplitudes (Fig. 2b); t$_1$ of 0.57 eV, as enhanced by the shorter Si-Si bond length due to the trimer distortion, and the t$_2$ and t$_3$ of 0.47 and 0.43 eV, respectively. 

\subsection*{Atomic and electronic structures of mobile phase defects}
Breaking translational symmetry by the trimer structure immediately leads to three degenerated ground states with fractionalized phase shifts of 0, 2$\pi$/3, and 4$\pi$/3 (Fig. 2b).
These ground states can be connected with a few different types of phase defects (or domain walls) as shown in Fig. 2c and Supplementary Fig. 7.
Only four of them are distinct, which we label with the distance between the neighboring Si atoms distorted.
Namely, a defect with a phase shift of 2$\pi$/3 (4$\pi$/3) corresponds to the $\times$2 (2a$_0$) or $\times$5 ($\times$1 or $\times$4) defects.
In order to identify detailed atomic and electronic structures of them, we performed DFT calculations with huge supercells (Supplementary Fig. 3). 
The results reveal that the $\times$4 structure is most stable in energetics [the formation energy of 0.092 ($\times$4), 0.124 ($\times$5), and 0.177 ($\times$2) eV/unit cell] (Supplementary Table I).
The simulated STM image for the $\times$4 structure reproduces fairly well the experimental ones discussed above, that is, the enhanced contrast at 0.1 V and the shifted protrusions at 1.0 V (Fig. 3a).
We also examined the other structure model proposed for the present system, the antiferromagnetic chain model \cite{erwi10}, but the phase defects could not be reproduced consistently (Supplementary Fig. 5).

The DFT (and also the tight-binding) calculations predict that the $\times$4 phase defect has its own electronic states within the band gap of the trimer chain as shown in Fig 3c.
The empty and filled states of the pristine 3a$_0$ Si chain are located at about +0.3 and -0.5 eV but the phase defect has its localized electronic state at around -0.2 eV. 
The localized in-gap state is clearly visualized in the STS map on a $\times$4 phase defect (Fig. 3c).
The phase shifts, atomic structures, and the in-gap electronic states detailed above converge convincingly to the topological soliton picture of the phase defects observed. 


Among four different types of phase defects (Fig. 2c), the $\times$4 defect was found to be most popular (Supplementary Fig. 1) in accord to the energetics calculated.
The $\times$1 defect is unstable to relax spontaneously into the $\times$4 defect.
The $\times$2 defect can also easily relax into the $\times$5 defect by simply recovering one distorted Si atom as shown in Fig. 2c.
The energy barrier of this process is as small as 0.01 eV (Supplementary Fig. 15).
Even the $\times$5 defect can transform into a more energetically favorable structure of two $\times$4 defects combined (called as $\times$4$\times$4) as shown in Fig. 2d.
The energy barrier is 0.06 eV being smaller than the hopping barrier of about 0.1 eV (Supplementary Fig. 15).

Indeed, we find quite a few $\times$4$\times$4 defect but rarely a $\times$5 defect  (Supplementary Fig. 1).
Note that the phase shifts themselves are preserved in these relaxation processes of the phase defects. 
The simulated STM image (Fig. 2f) of a $\times$4$\times$4 defect or a two soliton bound state is in good agreement with the experiment. 
Its electronic structure is similar to the isolated $\times$4 defect in both experiments and calculations (Supplementary Fig. 4) except for a small bonding-antibonding splitting (Supplementary Fig. 3).
The merging and splitting of two $\times$4 defects are hinted in the real time imaging (Fig. 1g and the supplementary movie1).

\subsection*{Topological nature and fractional charges}
As any other topological system, the topological nature of the present system is revealed by analyzing its band structure and edge states. 
The topological invariant of a trimer chain can be related to an effective higher dimensional (2D) bulk system theoretically \cite{krau12}. 
We construct such a 2D model by putting an adiabatic dimension and 
obtain the Chern numbers of (-1, 2, -1) for the three lowest energy bands (Supplementary Fig. 10) as predicted in previous theoretical studies \cite{mart19, ke16}.
The band gaps of the system contain five different edge states dictated by the topology (Fig. 2e), which match well the DFT calculations (Supplementary Fig. 9).
The major edge state of the C phase around 0.2 eV corresponds to the in-gap state observed in the experiment. 
A 2$\pi$/3 or 4$\pi$/3 fractional phase shift for an 1D electronic system guarantees fractionalized charges on corresponding solitons, while measuring the charge itself is a tremendous technical challenge - the tunneling occurs through not solitons themselves but electrons.
In theoretical aspects, we found that the 4$\pi$/3 phase-shift soliton has the fractionalized charges of +2e/3 (occupied) and -e/3 (empty) per spin and the 2$\pi$/3 phase-shift soliton has +e/3 (occupied) and -2e/3 (empty) per spin (Supplementary Fig. 12 and 13) \cite{conz20}.
The fractional charge is insensitive to detailed domain wall structures but depends only on the phase shift due to its topological origin.
For example, the fractional charge on a $\times$5 and a $\times$4$\times$4 defects is consistent {(Supplementary Fig. 13)}. 

\subsection*{Soliton motions}
We observe that the phase defects propagate at a higher temperature. 
At a 90 K, one seldom see the hopping of solitons, but, at 95 K, they exhibit several hoppings (by one 3a$_0$ unit cell of 1.16 nm) within a time window of 600 sec (Fig. 1e and 1f).
The hopping becomes more and more frequent with only a small change of the temperature as shown in Fig. 1f (Supplementary Movie1) and solitons become highly mobile already at 115 K. 
The drift velocity of the soliton at 100 K is measured as 0.10 nm/sec, which increase to 0.65 nm/sec at 115 K (Supplementary Fig. 14).
An estimation of Arrhenius-type diffusion velocity, D=D$_{0}$exp(-E$_{b}$/$k_{B}$T), gives the expectation of the velocity enhancement of 4.28 from 100 to 115 K (Supplementary Fig. 15a), which is roughly consistent with the observation. 
The turn-on of the soliton motion at around 100 K, related to the hopping barrier of a soliton 0.1 eV (Supplementary Fig. 15), seems consistent with the thermally induced disordering of the 3a$_0$ lattice, which was attributed to the generation of phase defects \cite{hafk20}. 
The real time images also clearly indicate that the soliton is immune to the defect scattering (it bounces back or jumps over the extrinsic defects, Fig. 1f) and the soliton-soliton scattering (they are reflected but prohibited to pass through, Fig. 1g, and Supplementary Movie1, and Supplementary Fig. 16). 
Of course, when the ground state structure of the Si chain is destroyed, for example, by impurity adsorption and increase of temperature, its edge modes, solitons, cannot be sustained. 

\subsection*{Generation of a single soliton}
A soliton is found to be reproducibly generated under the probe tip by the tunneling bias application at very low temperature. 
Figure 4a shows an atomically resolved atomic force microscopy (AFM) image of the surface at 4.4 K. 
In the AFM image, two undistorted Si atoms (blue atoms in the model of Fig. 2) of a trimer appear as dark contrast due to their closer distance to the tip. 
After the application of a single tunneling pulse at the location of the distorted Si atom (yellow circled in Fig. 4a), one can observe one trimer destroyed (Fig. 4b).
This transiently forms a $\times$6 chain in our structure model (Fig. 2) and relaxes into a $\times$5 soliton (Fig. 4c) and the phase shift of the neighboring trimers. 
This indicates the pair creation of $\times$1 and $\times$5 solitons with the former quickly moving out of the view frame to induce the phase shift. 
The soliton can also be erased by applying the same bias in a nearby site as shown in 
Figs. 4d-4f.
That is, the second soliton generated annihilates the first one. 
This switches back-and-forth the topological phase shift of a given trimer chain as shown in Figs. 4d-4f.  
That is, one can manipulate a single soliton and decode the topological phase information on each chain. 

\section*{Discussion}
A material realization of a fractionalized soliton has been elusive in an electronic system. 
Note that the popular dimer solitons has no electronic fractionalization due to the spin degeneracy. 
A close electronic example available is the phase defects in finite size artificial lattices based on a 2D surface state and adsorbates \cite{huda20}. 
However, this system only provides the static modulation of hopping amplitudes for an electronic orbital well away from the Fermi level to preclude the motion and the charge fractionalization.  
That is, these phase defects do not feature the dynamic nature, which is essential to a soliton.

The high mobility of the soliton observed directly here is notable since most of the solitons in previous works on solid surfaces are strongly pinned by defects or the strong interchain interaction \cite{lee19}. 
Mobile fractional solitons are contrasted with Majorana edge modes, for which an isolated mobile form has not been identified yet. 
The present solitons are further contrasted with Majorana modes and skyrmions by the fractionalized quanta associated. 
The soliton-soliton interaction glimpsed here as the formation of a soliton pair has an important implication in quantum information processing to secure an entangled state of solitons \cite{cold10,krup17,szum15,pend20}. 
The demonstration of the reproducible creation of an individual soliton here would open the door toward the manipulation of such information. 
Most of the essential ingredients for the exploitation of technological potentials of solitons in electronic systems are secured, such as the high mobility, the artificial generation/annihilation, the switchability \cite{kim17}, and the mutual interaction.

\section*{METHODS}
\textbf{Sample preparations.} The Si(553)-Au surface with a regular array of alternating Au chains and Si step-edge chains was fabricated by depositing Au of about 0.5 monolayer using a thermal evaporator onto a well cleaned Si(553) substrate at a temperature of 920 K in ultra high vacuum. The Si(553) substrate was cleaned by repeated flash heatings at 1500 K. The well ordered array was confirmed by low-energy-electron diffraction and STM images. 

\textbf{STM measurements.} The STM measurements were performed using a commercial low temperature STM apparatus in a ultra high vacuum chamber. The system was cooled down by liquid nitrogen while the temperature was carefully controlled by a built-in resistive heater. The measurement was done at various different temperatures between 78 and 125 K at various different biases. The tunneling current was typically fixed at 30 pA. The STS (dI/dV) measurement was performed using the standard lock-in technique with a lock-in modulation of 20 mV at 910 Hz and a tunneling current of 200 pA. For time-dependent topographic measurements for a short segment of a wire, the time resolution between successive scans is about a few sec.  

\textbf{AFM measurements.}
The imaging of the surface structure manipulation was performed by high-resolution noncontact atomic force microscopy (nc-AFM) and scanning tunneling microscopy (STM) under ultra high vacuum at 4.3 K using a commercial low temperature microscope (SPECS GmbH) with the simultaneous force-current detection capability. This combined system can effectively decouple the excitation source (tunneling current) and the local structural probe (AFM) so that the surface structure was investigated precisely by frequency shift ($\Delta f$) images while the tunneling current was limited to generate transient structures. The $\Delta f$ images were recorded at constant height. In the noncontact regime, $\Delta f$ increases in negative direction as the tip-sample distance gets closer, which means a more protruding structure appears darker. 

\textbf{Structure manipulation.} Step-edge Si trimer chains can be shifted reproducibly by exciting a local transient structure with the tunneling current injected. The local transient structure was generated using a bias pulse of +0.15 V on top of the trimer center at 4.3 K. After the sufficient time elapsed, the equilibrium state was achieved by the propagation of the soliton structure along the chain, which led to the phase shift in the trimer structure. We found that the initial excitation site, a trimer center, is prohibited from returning back to the sane trimer center, which forces a 1$a_{0}$ lateral shift of the original trimer structure. At a higher temperature than 4.3 K, a higher voltage is expected to induce the locally excited structure or solitons.

\textbf{DFT calculations.} DFT calculations were performed by using the Vienna $ab$ $initio$ simulation package \cite{kres96} within the generalized-gradient approximation (GGA) using the revised Perdew-Burke-Ernzerhof (PBEsol) functional \cite{perd08}.
The Si(553)-Au surface is modeled within periodic supercells with at least four bulklike Si layers and a vacuum spacing of about 12.8 {\AA}.
The bottom of the slab is passivated by H atom. 
We used a plane-wave basis with a kinetic energy cutoff of 312 eV and a 5$\times$2$\times$1 $k$-point mesh for the clean Si(553)-Au surface. All atoms but the bottom two Si layers held fixed at the bulk positions are relaxed until the residual force components were within 0.03 eV/{\AA}.
We employed large supercells for the defects (14 $\sim$ 17a$_0$) and their energetics is summarized in Supplementary Table I. It shows that its relative stability is sound, even though it is difficult to obtain the well-converged isolated energy of solitons due to huge and different sized supercells.
For the tight-binding model calculations, we used the PythTb package by Coh and Vanderbilt \cite{pythTB}.


\newcommand{\AP}[3]{Adv.\ Phys.\ {\bf #1}, #2 (#3)}
\newcommand{\CMS}[3]{Comput.\ Mater.\ Sci. \ {\bf #1}, #2 (#3)}
\newcommand{\PR}[3]{Phys.\ Rev.\ {\bf #1}, #2 (#3)}
\newcommand{\PRL}[3]{Phys.\ Rev.\ Lett.\ {\bf #1}, #2 (#3)}
\newcommand{\PRB}[3]{Phys.\ Rev.\ B\ {\bf #1}, #2 (#3)}
\newcommand{\PRA}[3]{Phys.\ Rev.\ A\ {\bf #1}, #2 (#3)}
\newcommand{\NA}[3]{Nature\ {\bf #1}, #2 (#3)}
\newcommand{\NAP}[3]{Nat.\ Phys.\ {\bf #1}, #2 (#3)}
\newcommand{\NAM}[3]{Nat.\ Mater.\ {\bf #1}, #2 (#3)}
\newcommand{\NAC}[3]{Nat.\ Commun.\ {\bf #1}, #2 (#3)}
\newcommand{\NAN}[3]{Nat.\ Nanotechnol.\ {\bf #1}, #2 (#3)}
\newcommand{\NAE}[3]{Nat.\ Electron.\ {\bf #1}, #2 (#3)}
\newcommand{\NARM}[3]{Nat.\ Rev.\ Mater.\ {\bf #1}, #2 (#3)}
\newcommand{\NL}[3]{Nano \ Lett.\ {\bf #1}, #2 (#3)}
\newcommand{\NT}[3]{Nanotechnology {\bf #1}, #2 (#3)}
\newcommand{\JP}[3]{J.\ Phys.\ {\bf #1}, #2 (#3)}
\newcommand{\JAP}[3]{J.\ Appl.\ Phys.\ {\bf #1}, #2 (#3)}
\newcommand{\JPSJ}[3]{J.\ Phys.\ Soc.\ Jpn.\ {\bf #1}, #2 (#3)}
\newcommand{\PNAS}[3]{Proc.\ Natl.\ Acad.\ Sci. {\bf #1}, #2 (#3)}
\newcommand{\PRSL}[3]{Proc.\ R.\ Soc.\ Lond. A {\bf #1}, #2 (#3)}
\newcommand{\PBC}[3]{Physica\ B+C\ {\bf #1}, #2 (#3)}
\newcommand{\PAC}[3]{Pure Appl.\ Chem. \ {\bf #1}, #2 (#3)}
\newcommand{\SCI}[3]{Science\ {\bf #1}, #2 (#3)}
\newcommand{\SCA}[3]{Sci.\  Adv.\ {\bf #1}, #2 (#3)}
\newcommand{\SR}[3]{Sci. \ Rep. \ {\bf #1}, #2 (#3)}
\newcommand{\RPP}[3]{Rep.\ Prog.\ Phys. \ {\bf #1}, #2 (#3)}
\newcommand{\NPJQ}[3]{npj \ Quantum \ Materials \ {\bf #1}, #2 (#3)}
\newcommand{\RMP}[3]{Rev. \ Mod. \ Phys. \ {\bf #1}, #2 (#3)}

\begin{figure*}[htp] 
\centering{ \includegraphics[width=16.0 cm]{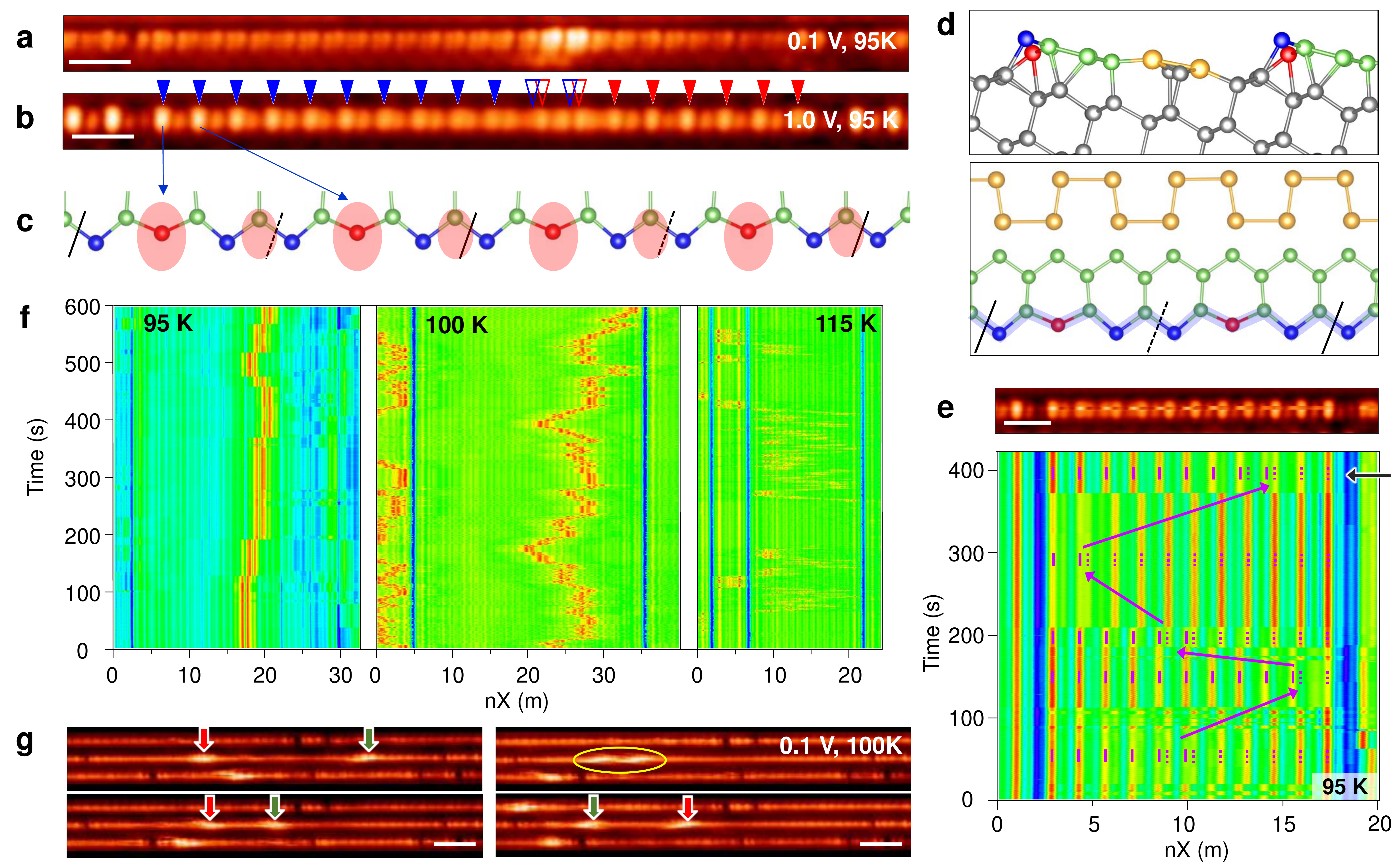} }
\caption{ \label{fig1}
{\bf Mobile phase defects on Si chains of a Si(553)-Au surface.}
{\bf a} and {\bf b} STM images of a phase-shift defect along a Si step-edge chain in its CDW state at 95 K (sample bias voltage $V_s$ of 0.1 and 1.0 V; scale bar of 2 nm).
Blue (red) triangles denote the $\times$3-period CDW on the left (right) domain.
Empty triangles indicate the phase mismatch.
{\bf c} Schematic atomic structure of a Si step-edge chain in the CDW state.
The ovals denote the protrusions observed in STM with a bias of 1.0 V {\bf b}.
{\bf d} Atomic structure (side and top view) model of the periodically distorted CDW phase \cite{brau18}.
Yellow, green, and gray balls represent Au, top-layer Si, and bulk Si atoms, respectively.
Red and blue balls represent the distorted and undistorted Si atoms at the step edge, respectively.
{\bf e} A continuous real time measurement of the STM profiles ($V_s$ = 1.0 V) for the same chain segment for a time interval of about 400 sec.  
Purple lines indicates the 3a$_0$ CDW and purple arrows the phase shift defect.
Top panel is a STM images ($V_s$ = 1.0 V) of a Si chain showing the lateral shift of the CDW related to the hoppings of a phase defect.
{\bf f} Similar real time STM profile measurements at three different temperatures but with a low bias of 0.1 eV, where the phase defect is images with a strong contrast (red colored in the profiles). 
{\bf g} Snapshot STM images ($V_s$ = 0.1 V) of the mobile phase defects at 100 K (scale bar = 4 nm). See also the Supplementary Movie 1.
The arrows indicate the two particular phase defects and the ellipse indicates the temporary pairing of them. 
}
\end{figure*}

\begin{figure*}[htp] 
\centering{ \includegraphics[width=16.0 cm]{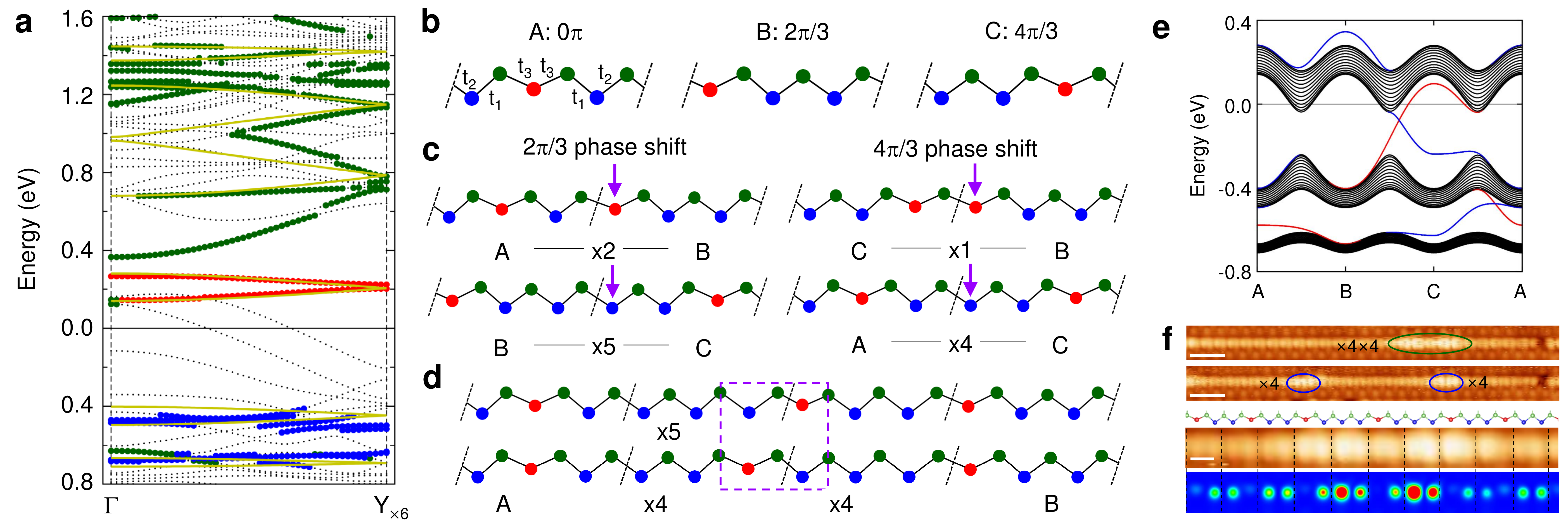} }
\caption{ \label{fig2}
{\bf Electronic structure of the Si(553)-Au surface in its CDW state.}
{\bf a} Band dispersion along the step edge as calculated by DFT.
Red, blue, and green circles denote the localized states more than 15 \% at outer distorted, undistorted Si atom and inner Si atoms of zigzag chain, respectively. 
Yellow lines denote the tight-binding band of the zigzag Si chain.
{\bf b} Schematics of three translationally degenerate phases (A, B, and C) for the distorted Si chain which corresponds to the step-edge zigzag Si chain marked by the blue shade in Fig. 1{\bf d}.
{\bf c} Four distinct phase boundary (defect) structures.
{\bf d} Comparison between a $\times$5 and a $\times$4$\times$4 structure with the same overall phase shift.
{\bf e} Energy spectrum for the adiabatic evolution of open boundary trimer chain (15 unit cell). The tight-binding parameters were taken for the phase evolution of A\textrightarrow B\textrightarrow C\textrightarrow A. The black lines denote the eigenstates of the open boundary chain and red and blue lines correspond to localized states at right and left edges, respectively. 
{\bf f} Two successive STM images of a Si chain with two $\times$4 defects, paired or separated ($V_s$ = 1.0 V at 95 K, scale bar = 2 nm) and enlarged experimental and simulated STM images of a $\times$4$\times$4 defect ($V_s$ = 0.1 V, scale bar = 0.5 nm).
}
\end{figure*}

\begin{figure*}[htp] 
\centering{ \includegraphics[width=16.0 cm]{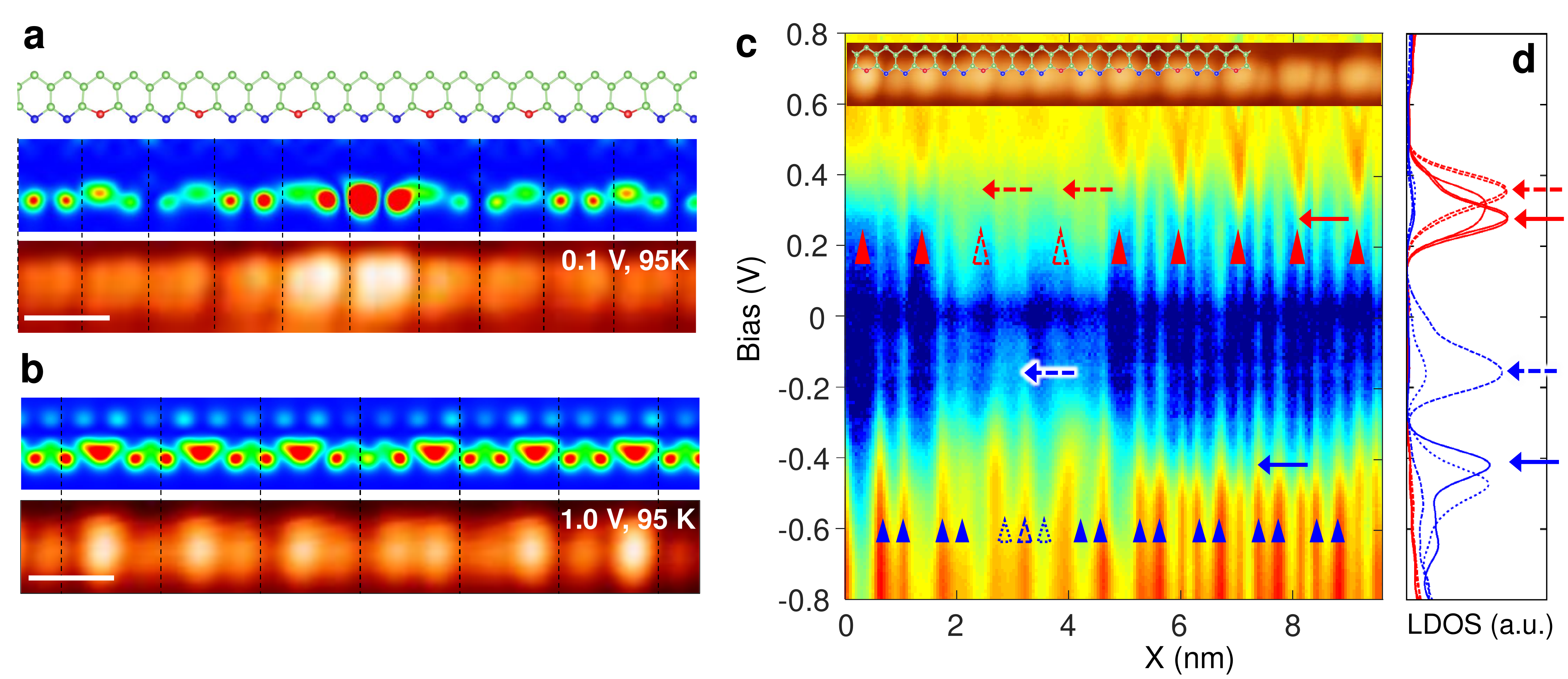} }
\caption{ \label{fig3}
{\bf Atomic and electronic structures of a $\times$4 soliton.}
{\bf a} and {\bf b} Simulated and experimental STM images at 0.1 V and 1.0 V, respectively (scale bar = 1 nm) for a $\times$4 phase defect.
{\bf c} STS (dI/dV) line profile along a Si chain taken at 95 K including a $\times$4 phase defect.
The arrows indicate the peak position of the calculated LDOS in {\bf d} at the pristine chain (solid lines) and on the defect (dashed lines).
Blue (red) lines denote the localized states at undistorted (distorted) Si atoms.
Long (short) blue dashed line denote the central (outer) Si atom of the defect.
}
\end{figure*}

\begin{figure*}[htp] 
\centering{ \includegraphics[width=16.0 cm]{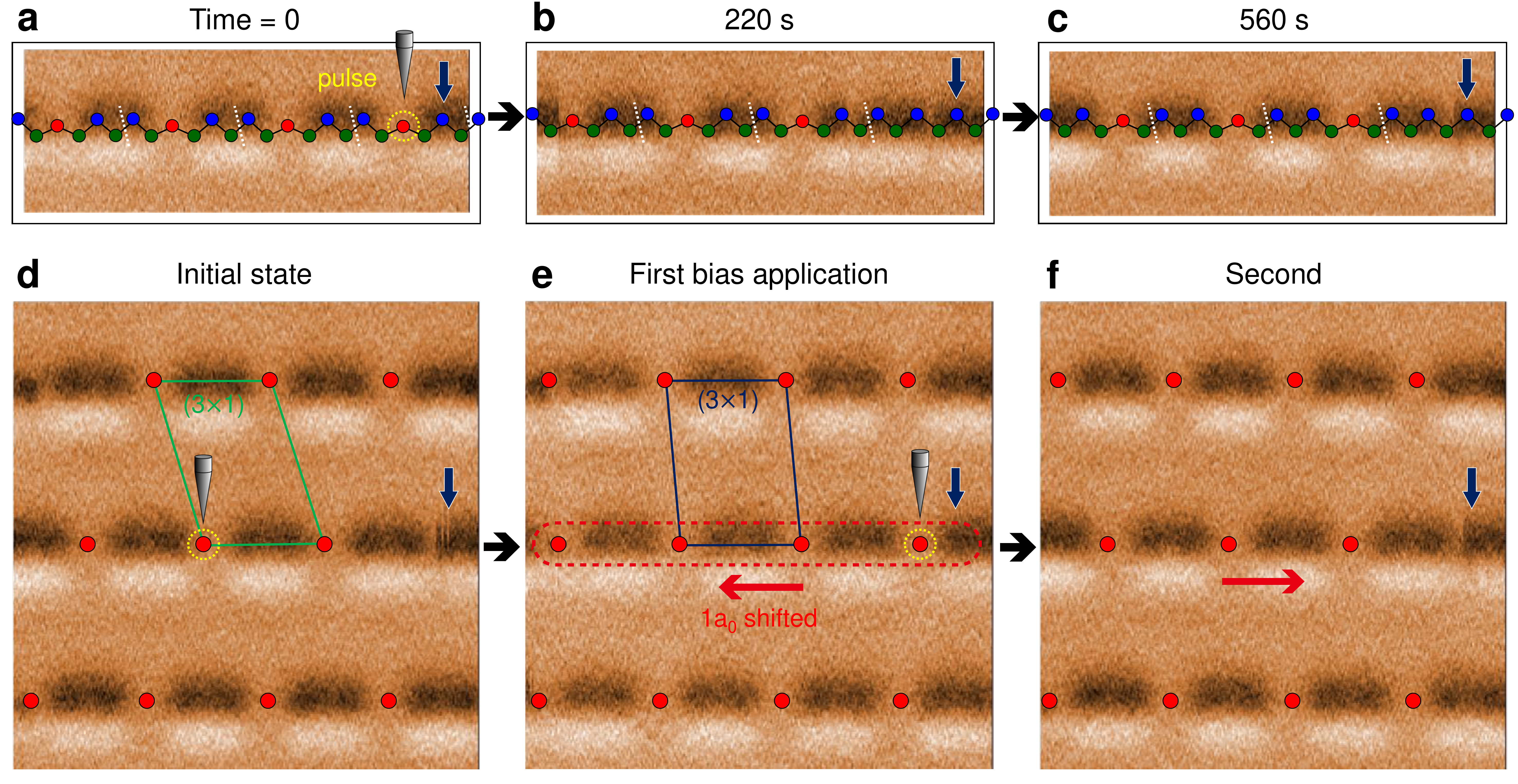} }
\caption{ \label{fig4}
{\bf Creation of a single soliton.}
Frequency-shift non-contact atomic force microscopy image on a Si trimer chain just before {\bf a} and after {\bf b}-{\bf c} the injection of the tunneling pulse from the metallic probe tip. The excitation and relaxation of the trimer chain is imaged in atomic scale. The atom indicated by the arrow traps the soliton created but is not altered by the tunneling pulse and is thought to be pinned by a defect or an impurity nearby. {\bf d}-{\bf f} Switching of the selected mid trimer chain structure using on-site bias pulse. Positions of trimer center atoms and corresponding single unit cell are indicated for clear comparisons between transitions.
}

\end{figure*}
\end{document}